# Exploring Geometric Shapes with Touch

Thomas Pietrzak[1], Andrew Crossan[2], Stephen A. Brewster[2], Benoît Martin[1], Isabelle Pecci[1]

[1]Université Paul Verlaine – Metz, UFR MIM, île du Saulcy
57006 Metz, France
[2]University of Glasgow, Department of Computing Science
18 Lilybank Gardens, Glasgow, UK.
{Thomas.pietrzak, benoit.martin, isabelle.pecci}@univ-metz.fr
{ac, stephen}@dcs.gla.ac.uk

**Abstract.** We propose a new technique to help users to explore geometric shapes without vision. This technique is based on a guidance using directional cues with a pin array. This is an alternative to the usual technique that consists of raising the pins corresponding to dark pixels around the cursor. In this paper we compare the exploration of geometric shapes with our new technique in unimanual and bimanual conditions. The users made fewer errors in unimanual condition than in bimanual condition. However they did not explore the shapes more quickly and there was no difference in confidence in their answer.

## 1 Introduction

Many subjects at school rely on structured data such as schematics. These are essentially visual representations, which are difficult to interpret for children with visual impairment. Children with some residual vision may be able to use magnified schematics, but those who are blind or have little residual vision have to use another sense. Raised paper is widely used for this purpose. Children can explore raised paper schematics by running their fingers over the paper to feel the raised sections of the image. However this technique provides a static representation of the schematic and cannot easily benefit from the advantages of a computer-based system. Large Braille displays could potentially provide similar functionality for a computer-based system, but are prohibitively expensive for most situations. The goal of our research is to identify techniques that can provide accessible representations of structured data such as shapes or schematics, and both take advantage of the benefits of the technology-based approaches without being too expensive. The VTPlayer mouse (figure 1) is an example of such a display, which has two 4x4 pin arrays on the top. It has previously been used to display tactile icons, which appeared to be a good way of conveying information [4]. We are interested in two aspects concerning the display of schematics with touch. The first one is the way the information is presented, and the second one is the way the schematic is explored. In this paper we propose a new presentation method, and will compare two exploration methods for it.

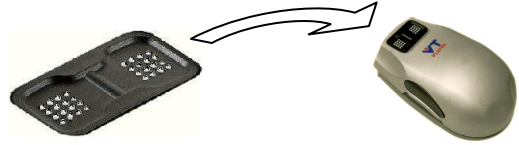

**Fig. 1.** VTPlayer mouse, with two 4×4 pins matrices on the top

## 2  Presentation and exploration methods

The common method to display schematics with pin arrays is a visual translation method that is a simple mapping of dark pixels into raised pins. It has been shown to be an inefficient method [2], unless it is enhanced by a guidance system [1,3].

The new method we propose consists of guiding users along the shape in order to help them explore it completely. It is based on a vectorial representation of the shape to explore. The shape is divided into line segments, which the user is guided around using directional Tactons. Tactons are structured tactile cues that convey information [4]. We use Tactons presented through one 4x4 pin array of a VTPlayer, which has been shown to be able to represent efficiently 8 directions (figure 2) [5]. A preliminary study of shape exploration with pin array Tactons in unimanual condition [6] shows that users were able to explore shapes with this technique. However they only provided directional information. The next step would be to enhance this system with more information. According to initial tests on Tactons, we can transmit successfully several pieces of information by varying several parameters on one Tacton [5]. For example, dynamic "blinking" Tactons are made with two individual animated frames. The Tacton is displayed by alternating a frame with a pattern made with raised pins and another frame with no raised pins with both frames being displayed the same duration.

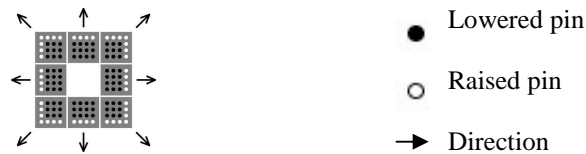

**Fig. 2.** Patterns used to encode direction

Here we can display one piece of information through the pattern and another through the blink speed. We still use the pattern to encode the direction of the next vertex, and we use the blink speed to encode the distance to the next vertex. Initial studies varying blink speed showed that 3 blink speed values can be distinguished by participants with a high level of accuracy [5]. Here, we map increasing blink speed to decreasing distance to the start of the next line segment. Finally, we use a simple binary mapping on the second pin array of the VTPlayer which rests under the user's middle finger to encode whether the user's cursor is on the shape or outside the shape. The segments have a thickness, fixed after pilot studies. The opposite vertex is called the target. It is represented by a half-circle whose radius is equal to the segment's thickness (figure 3). The user is guided from one vertex to the next one in a loop, and can move around the shape as many times as is required. The Tacton is displayed under the index finger, and its characteristics depend on the user's position with respect to the current segment's position. If the user is on the segment, the Tacton's pattern aims at the target. Otherwise, it aims at the nearest point on the segment.

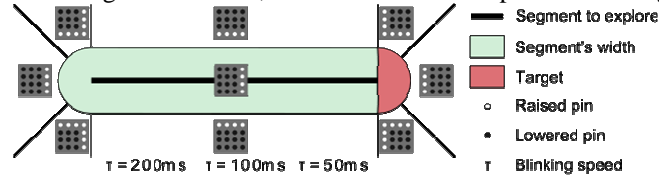

**Fig. 3.** Guidance on a segment

We explore two conditions: unimanual and bimanual. In the unimanual condition, users both navigates with the VT Player and feels the Tactons through their dominant hand, In the bimanual conditions, users navigate using a stylus in their dominant hand, while feeling the stimulus on their non-dominant hand resting on a static VT Player (as in [1,3]). The advantage of the first one is that the kinesthetic sensation provided by the movement of the pointing device comes from the same part of the body as the tactile stimulation. The advantage of the second method is that it could provide an alternative solution for blind users, for whom previous work has suggested that the use of the mouse can be problematic [2]. The use of an absolute referential provided by a tablet could resolve these problems.

## 3  Experiment

In this preliminary experiment we will search for the best exploration methods for our new presentation method. The task was to identify a series of shapes as quickly and accurately as possible. Eight sighted but blindfolded users took part of this experiment. They were 26 to 28 years old, and none of them suffers from a known tactile perception problem. All of them were regular computer users, but none had significant experience of using tablets. Half of the users explored the shapes with the VTPlayer, and the others with the tablet. All participants explored one training shape until they felt comfortable with the system, and 10 shapes (including a square, a triangle, a rectangle and simple polygons) they had 3 minutes to recognize. Response times as well as a confidence rate (on a scale of 1 to 7) were recorded for each shape. We make the hypothesis that users will make less errors, answer more quickly and will be more confident in their answers when exploring with the VTPlayer mouse rather than with the tablet.

We notice that the users who used the tablet made more errors (8/40) than the others who used the VTPlayer mouse to explore (4/40). Users who explored the shapes in unimanual condition gave a mean confidence of 5.85/7 (sd=1.28), compared to 5.6 (sd=1.62) for users who explored in bimanual condition. A Wilcoxon rank sum test does not show any significant difference (W=878, p=0.43). Concerning answer times, users who explored in the unimanual condition answered with a mean of 95.95s

(sd=42.77s) compared to a mean of 93.45s (sd=52.54s) in the bimanual condition. Once again a Wilcoxon rank sum test doesn't show any difference (W=881.5, p=0.43).

According to these results we cannot accept our hypothesis about the error rate, the confidence rate and the exploration time. Indeed users made few errors with the VTPlayer and with the tablet. Moreover the exploration method did not have any impact on the answer times nor on the confidence of the users on their answers. However, some users reported it was disturbing to explore with one hand and feel the shape with the other one. These results are encouraging for the Tacton representation method as users were able to achieve a low error rate in both mouse and tablet exploration modes with very little training. We must be aware however that results might differ for visually impaired users

## 4   Conclusion

We proposed a new presentation method for geometrical shapes. In this preliminary study we have tested the difference between two exploration methods with this new presentation method. Our results show that sighted users could use a Tacton representation to successfully explore shapes non visually. There were no significant differences in time or confidence when exploring shapes in unimanual condition or in bimanual conditions. The similar performance of these exploration methods encourages us to concentrate on exploration with a tablet as an input as this may be more suitable for visually impaired users due to the absolute positioning method used by this device. In future studies we will compare the efficiency of the new presentation method with the traditional tactile presentation method that consists of raising pins corresponding to dark pixels around the cursor, with sighted and visually impaired users.